\documentclass[runningheads]{llncs}
\usepackage{cite}
\usepackage{amsmath,amssymb,amsfonts}
\usepackage{algorithmic,algorithm}
\usepackage{graphicx}
\usepackage{textcomp}
\usepackage{xcolor}
\usepackage{tikz}
\usepackage{pgfplots}
\usepackage{pgfplotstable}
\usepackage{bm}
\usepackage{supertabular}
\usepackage{caption}
\usepackage{subcaption}
\usepackage{url}
\usepackage{booktabs} 
\usepackage{multirow}
\usepackage{tabularx} 
\usepackage{pgfplots}
\usepackage{array}     
\usepackage{tabularx}   
\usepackage{comment}
\usepackage{makecell} 
\usepackage{tikz}
\usetikzlibrary{arrows.meta,positioning,calc}
\usepackage{xcolor}
\usepgfplotslibrary{groupplots}
\pgfplotsset{compat=1.18}

\definecolor{oiVermilion}{HTML}{D55E00} 
\definecolor{cmtGreen}{HTML}{009E73}     
\definecolor{rb1}{RGB}{255,0,0}     
\definecolor{rb2}{RGB}{191,0,64}
\definecolor{rb3}{RGB}{128,0,128}
\definecolor{rb4}{RGB}{64,0,191}
\definecolor{rb5}{RGB}{0,0,255}     

\tikzset{
  rlnc/.style={solid, line width=1.2pt, mark=+, color=black, mark options={scale=1.2, solid}},
  twodrs/.style={line width=1.2pt, color=oiVermilion,
                 mark=star, mark options={solid, scale=1.2}},
  cmt/.style={solid, line width=1.2pt, color=cmtGreen,
              mark=diamond, mark options={fill=white, solid, scale=1.2}},
  uncoded/.style={line width=1.2pt, color=gray,
              mark=o, mark options={solid, scale=1.0}},
}
\pgfplotstableread[col sep=comma]{plot_data/soundness_failure_probability.csv}\SFP
\pgfplotstableread[col sep=comma]{plot_data/da_challenge_response_size.csv}\DARESP
\pgfplotstableread[col sep=comma]{plot_data/da_commitment_size.csv}\DACOMM
\pgfplotstableread[col sep=comma]{plot_data/total_download_cost.csv}\DATOTAL
\pgfplotstableread[col sep=comma]{plot_data/storage_cost.csv}\DASTORAGE
\pgfplotstableread[col sep=comma]{plot_data/comp_cost.csv}\DACOMP

\begin{document}

\title{
From Indexing to Coding: A New Paradigm for Data Availability Sampling
}
%
%
\author{Moritz Grundei \and
Vipindev Adat Vasudevan \and
Kishori Konwar \and
Muriel M\'edard
}

\authorrunning{M. Grundei, V. A. Vasudevan, K. Konwar, M. M\'edard}

\institute{Optimum, Cambridge, MA, USA \\
\email{\{moritz,vipindev,kkonwar,mmedard\}@getoptimum.xyz}}

\maketitle             
\begin{abstract}
The data availability problem is a central challenge in blockchain systems and lies at the core of the accessibility and scalability issues faced by platforms such as Ethereum. Modern solutions employ several approaches, with data availability sampling (DAS) being the most self-sufficient and minimalistic in its security assumptions. Existing DAS methods typically form cryptographic commitments on codewords of fixed-rate erasure codes, which restrict light nodes to sampling from a predetermined set of coded symbols.

In this paper, we introduce a new approach to DAS that modularizes the coding and commitment process by committing to the uncoded data while performing sampling through on-the-fly coding. The resulting samples are significantly more expressive, enabling light nodes to obtain, in concrete implementations, up to multiple orders of magnitude stronger assurances of data availability than from sampling pre-committed symbols from a fixed-rate redundancy code as done in established DAS schemes using Reed Solomon or low density parity check codes. We present a concrete protocol that realizes this paradigm using random linear network coding (RLNC).

\end{abstract}
\section{Introduction}

\begin{figure}
    \centering
    \includegraphics[width=0.85\textwidth]{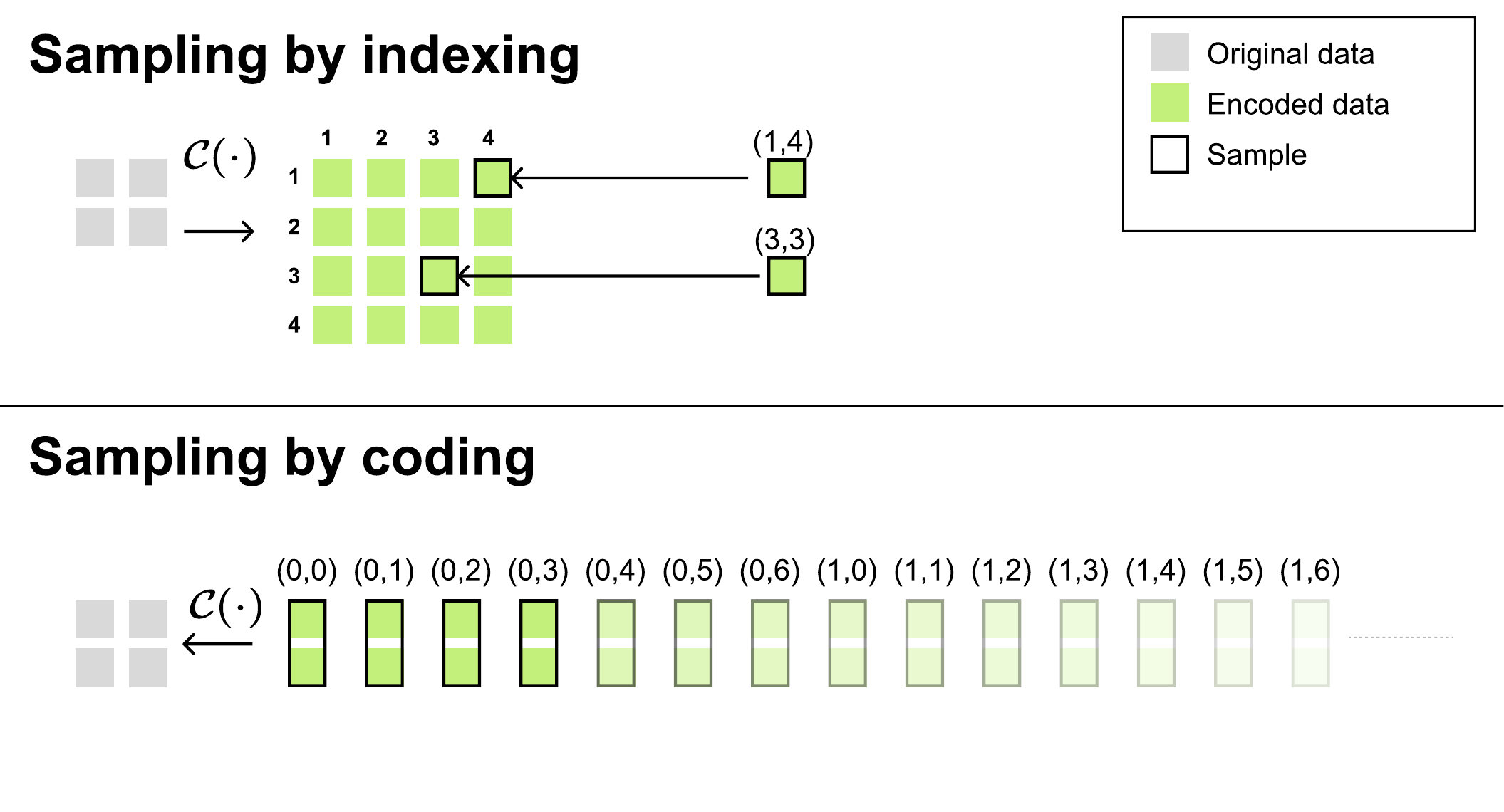}
    \caption{Two sampling paradigms. Top: Sampling by indexing, where samples are generated by returning fixed rate coded data from an array of coded data. Bottom: Sampling by coding, where a dynamic coding operation is performed during the sampling process.}
    \label{fig:static_sampling_vs_dynamic_sampling}
\end{figure}

\par
Blockchains such as Ethereum can be modeled as distributed state machines that rely on full replication of a shared state. They function by disseminating transaction data throughout the network, aggregating it into sequentially linked blocks, and applying this data to update the globally shared state. In recent years, scalability and accessibility have become central challenges: As network throughput has increased, the hardware requirements for operating a full node have also risen. To address this, Ethereum and similar blockchains support light nodes alongside full nodes. Light nodes do not store or download the full transaction history; instead, they maintain custody of block headers, which act as compact references to the data contained in each block. While this design allows light nodes to verify the inclusion of individual transactions, it does not provide them with any assurances regarding the integrity or even the availability of data to the system of nodes.

\par
The problem of verifying that the data contained in a block is available to a system of nodes is known as the data availability problem. Several approaches exist to provide availability guarantees, but one of the most minimalistic and self-sufficient methods is data availability sampling (DAS) \cite{2DRSDataAvailability, hall2023foundations}. In DAS, a light node challenges nodes that claim that the data is available by requesting small, randomly chosen portions of the block data. Each successfully answered challenge provides the light node with probabilistic assurance that the full data is indeed available. To prevent an adversarial node from selectively withholding only a small fraction of the data, DAS protocols operate over coded data rather than raw block data, ensuring that even partial withholding can be detected \cite{2DRSDataAvailability, CodedMerkleTree, hall2023foundations, DAMultiplicity}.

\par
Previous approaches combine fixed-rate erasure-coded data with a cryptographic commitment layered on top \cite{hall2023foundations, 2DRSDataAvailability, CodedMerkleTree, DAMultiplicity, YuEtAl-DeterminingDataAvailability-2022, YuEtAl-VerifyingRemotelyStoredData-2022}, which serves as a reference for light nodes to verify the integrity of the sampled data shares. In this setting, a light node performs sampling with an index sampler, selecting coded symbols at specified positions from the underlying array of coded data. Since the commitment was generated for this single coded representation, it fixes the coded representation and significantly reduces the degrees of freedom for the sampling operation. Further, the size of the sampling space is inherently constrained by the storage capacity of the participating nodes, either producing the data or claiming it available.

\par
In this paper, we propose a new, modular paradigm for DAS that decouples the cryptographic commitment from the coding process. The key idea is to employ a sampling method that allows a claimer to generate coded data on the fly, while using a cryptographic commitment to the uncoded data. This design offers several advantages.
First, it  separates the commitment scheme from the coding of the underlying data offering more modularity. Second, the sampling space is no longer constrained by the storage capacity of the individual nodes, which makes the set of possible challenges richer compared to sampling from pre-committed coded data. Finally, by generating samples dynamically at the claimer, we trade additional computation at the claimer for reduced storage requirements and lower download costs for the light node, since the resulting samples are more powerful in their impact on certainty about data availability than those derived from pre-committed coded data.

\par
Further, we demonstrate a concrete implementation of this new approach to DAS using random linear network coding (RLNC) \cite{ho2006random}, a class of rateless erasure codes. RLNC is particularly well suited to our modular paradigm because it allows coded samples to be generated flexibly and efficiently on demand and integrates naturally with a  commitment scheme on the uncoded data using homomorphic vector commitments like Pedersen commitments. Within this framework, a single RLNC sample can provide the same certainty about availability as approximately 73 samples from a commonly used $r = 0.25$ two-dimensional Reed–Solomon (RS) code and 156 samples from the low density parity check (LDPC) codes employed in \cite{CodedMerkleTree, LDPCForCMT, PeerDAS}. In our simulations, this translates into (i) a reduction in download costs for light nodes by a factors of up to $\sim 12 - 195$ compared to schemes that use pre-committed codewords of 2D-RS codes or LDPCs, while (ii) incurring no storage overhead at the claimer node, in contrast to the $4\times$ storage overhead of the original data at claimer nodes holding redundancy coded data.

\section{Notation}

We use the following notation throughout the paper. Scalars are denoted by lowercase letters (e.g., $a$), vectors by bold lowercase letters (e.g., $\mathbf{c}$), and matrices by bold uppercase letters (e.g., $\mathbf{A}$). For a matrix $\mathbf{A}$, we let $\mathbf{a}_i$ denote its $i$-th column and $\mathbf{a}^{(i)}$ its $i$-th row. The field of coding coefficients is denoted by $\mathbb{F}$, and, depending on the context, $|\mathbb{F}|$ may either refer to the cardinality (the number of elements in the field) or the size of a binary representation of any element in the field. A set of $n$ elements is denoted by $\{a_i\}_{i=1}^n$, while a binding but not hiding Pedersen vector commitment to a vector $\mathbf{a} \in \mathbb{F}^n$ with respect to the generator basis $\{g_1, \dots, g_n\}$ is denoted by$[[\mathbf{a}]]_g = \sum_{i=1}^{n} a_i g_i \, .$
Finally, an inner product argument for vectors $\mathbf{a}$ and $\mathbf{b}$ is written as $\langle\!\langle \mathbf{a}, \mathbf{b} \rangle\!\rangle$.

\section{Sampling from Coded Data}

\begin{figure}
    \centering
    \includegraphics[width=0.85\linewidth]{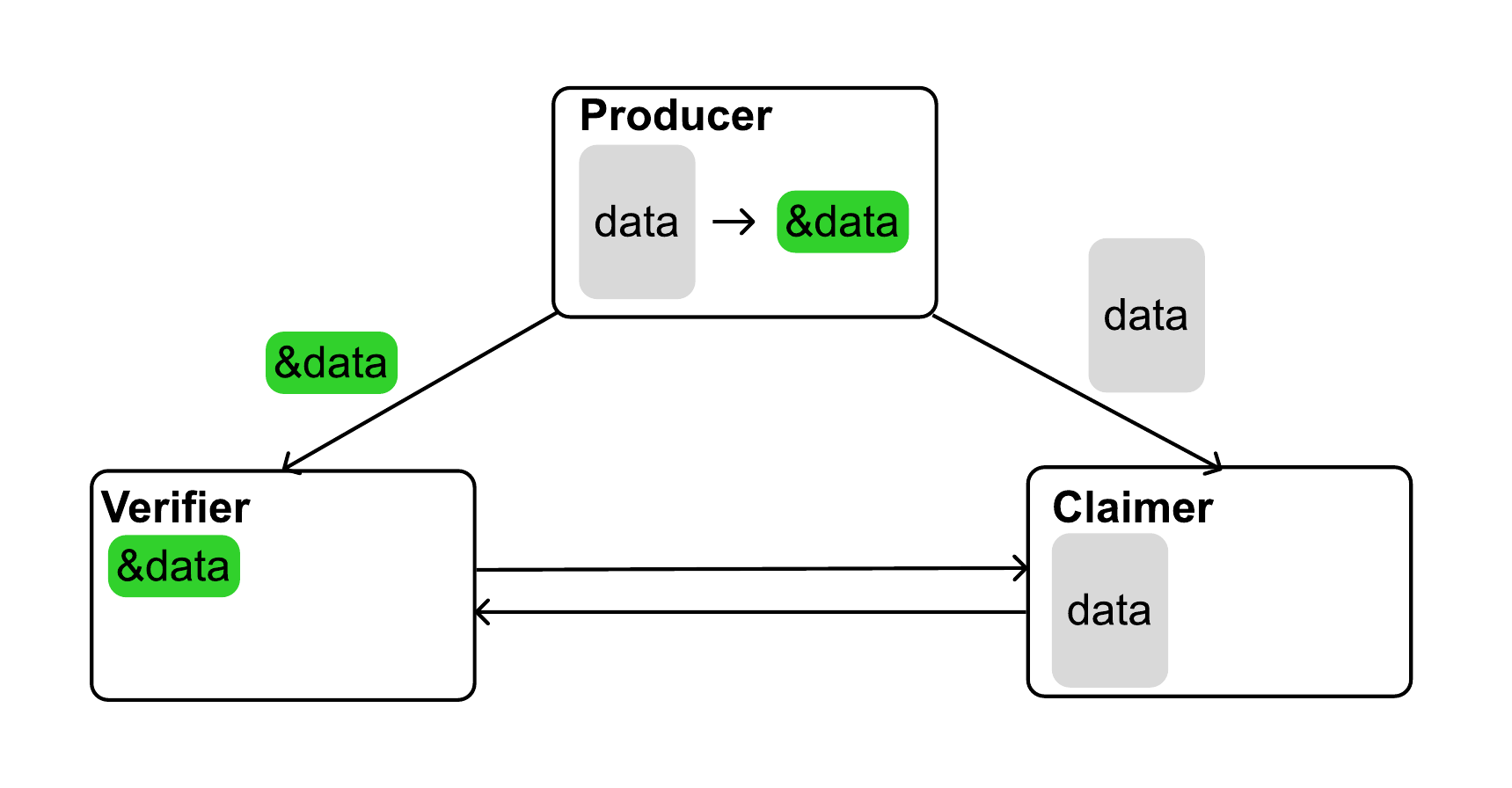}
    \caption{System actors in the data availability problem: The verifier tries to ascertain the availability of the data corresponding to a data reference obtained from producer through requesting coded samples from a claimer node.}
    \label{fig:DA_setup}
\end{figure}

\subsection{System Model}
\label{sec:model}

\par
In a trustless system of nodes where resource constrained nodes want to gain certainty about data being available to the system, Fig. \ref{fig:DA_setup} displays the three system actors:
A data producer creates a succinct cryptographic data reference to the data entity (or a coded version thereof) and broadcasts this reference to participating nodes in the system (e.g. by putting it into the header of a block in a blockchain). A verifier node with limited bandwidth and / or storage resources (e.g. an Ethereum light node) then queries one or several claimer nodes that claim that the data is available to the network. The claimer node responds with a coded sample of the underlying data, accompanied by a membership proof that can be used to verify that the sampled data was derived from the data entity. For the purposes of this work, we additionally adopt a simplifying assumption regarding the interaction between verifier and claimer nodes, as well as the sampling strategy, namely the strong model for data availability~\cite{neu2022data}. In particular, we assume non-adaptivity of claimer nodes: whenever a claimer attempts to deceive verifiers about data availability by revealing only a subset of coded chunks, it must decide which shares to publish before the sampling process begins.

\par
For schemes that employ layered commitments on coded data that are not code binding (e.g., \cite{2DRSDataAvailability, CodedMerkleTree}), an additional actor, an auditor node, is required. The role of the auditor is to validate the responses provided by claimer nodes. If a response is incorrectly coded the auditor can generate and transmit a fraud proof, thereby alerting the verifier to the inconsistency.

\subsection{Sampling Coded Data: Two Paradigms}

\begin{figure}
  \centering
  \begin{tikzpicture}
    \begin{axis}[
      ymode=log, ymin=1e-10, ymax=1,
      xmode=log,
      xtick={1,10,100},
      xticklabels={1,10,100},
      xlabel={Number of samples}, ylabel={$P_{\text{FN}}$},
      xmin=1, xmax=100, 
      grid=major, major grid style={gray!30,dashed},
      legend columns=5,
      legend style={
        at={(0.5,1.10)}, anchor=south,
        draw=black, fill=white, rounded corners=2pt,
        /tikz/every even column/.append style={column sep=0.8em}
      },
      label style = {font=\large},
      width=0.8\textwidth,
      height= 6cm,
      unbounded coords=discard
    ]
      \addplot+[rlnc, mark=+, color=rb3]  table[x index=0, y index=2]{\SFP}; \addlegendentry{RLNC}
      \addplot+[twodrs] table[x index=0, y index=4]{\SFP}; \addlegendentry{2D-RS}
      \addplot+[cmt]   table[x index=0, y index=5]{\SFP}; \addlegendentry{LDPC}
    \end{axis}
  \end{tikzpicture}
  \caption{Probability that withholding a sufficient amount of samples to make the underlying data entity undecodable goes undetected (false negative probability $P_{FN}$) when (i) sampling by indexing code words of fixed-rate $r = 0.25$ 2D-RS or LDPC codes, versus (ii) sampling by coding with RLNC using a field of cardinality $2^{16}$ (2B).}
  \label{fig:prob_soundness_failure}
\end{figure}

\par
When sampling from a coded representation of a data entity, two paradigms must be distinguished:
\begin{enumerate}
    \item \textbf{Sampling by indexing}: Sampling is performed from a static, precomputed set of coded pieces. The verifier specifies which pieces to retrieve, for example by providing indices to an array of coded symbols. Consequently, the sampling space is bounded by the size of the pre-stored data at the claimer which only forwards the coded symbols produced by the producer on request from the verifier.
    \item \textbf{Sampling by coding}: In this setup, the verifier provides an input parameter to a coding operation $\mathcal{C}(\text{data}, \cdot)$, which prompts the claimer to compute the response dynamically rather than retrieving it from a precomputed array, as in the sampling by indexing paradigm. The size of the sampling is thus not determined by the size of the encoded data stored at the claimer node, but instead follows the cardinality of the input to the coding function.

\end{enumerate}

\par
Figure~\ref{fig:static_sampling_vs_dynamic_sampling} illustrates the two sampling paradigms. In the upper half, sampling is performed from precomputed coded data arranged as a $4 \times 4$ array of 16 coded symbols, exemplifying the sampling by indexing approach. The verifier specifies coordinates (e.g., $(1,4)$) to retrieve the corresponding symbol. In this case, the size of the sampling space is fixed at 16, matching the number of coded symbols stored at the claimer node.

\par
The lower half of the figure illustrates sampling by coding. In this paradigm, the coded symbol returned to the verifier is computed dynamically from the underlying data together with an input tuple (e.g., $(0,1)$). The size of the sampling space is determined by the cardinality of the field from which the input tuple is drawn. In the example shown, the tuple consists of elements from the prime field $GF(7)$, resulting in a sampling space of size $7^2 = 49$. This additional flexibility in specifying a coded sample, when combined with a suitable coding operation, enables the verifier to derive a substantially stronger estimate of the probability that the underlying data is actually available as is shown next in a concrete embodiment of this paradigm with RLNC.

\par
This example illustrates an important trade-off: sampling by indexing a coded data representation is storage-intensive but computationally cheap during sampling, whereas sampling by coding minimizes storage requirements at the cost of increased computational complexity during sampling. As we will see in the following example, sampling by coding, when applied with a suitable coding scheme, yields much stronger guarantees for verifier nodes probing claimer nodes suspected of withholding data from the network.

\subsubsection{Example Paradigm 1: Fixed Rate Redundancy Codes}

\par
Prominent DAS schemes combine fixed-rate linear redundancy codes with index sampling \cite{2DRSDataAvailability, CodedMerkleTree, hall2023foundations, PeerDAS}. Given $k$ data symbols to be checked for availability, the producer encodes the $k$ symbols into $n$ coded symbols, which the claimer then forwards to verifiers on request. The code rate is $r = k/n$.

\par
For a malicious claimer to prevent decoding, they must withhold enough coded symbols to prevent reconstruction from the set of coded pieces emitted. In this context, the undecodability ratio $\alpha$ denotes the minimum fraction of coded symbols that must be hidden to keep the data unrecoverable from the sampled pieces. Intuitively, the smaller $\alpha$ is, the more samples a verifier must request to gain high assurance of data availability, since the adversary can reveal a larger portion of symbols while still preventing reconstruction.

\par
Several embodiments of this paradigm appear in the literature \cite{2DRSDataAvailability, hall2023foundations, DAMultiplicity, CodedMerkleTree, LDPCForCMT, PeerDAS}. In \cite{2DRSDataAvailability}, a two dimensional Reed Solomon (2D-RS) approach is described whereby $k$ chunks are arranged into a $\sqrt{k}\!\times\!\sqrt{k}$ matrix and subsequently encoded row- and column-wise with a $r=0.5$ RS code. To derive data references, each row and column of the encoded matrix is treated as the leaf set of a Merkle tree, and the resulting Merkle roots are placed in the block header. Variants also pair 2D-RS with polynomial commitment schemes such as KZG, which reduce both commitment and sampling sizes at the expense of requiring a trusted setup (e.g., a KZG ceremony) \cite{DAMultiplicity, PeerDAS}.
The coded Merkle tree (CMT) construction in \cite{CodedMerkleTree} proposes a layered data tree with LDPC-coded layers. This construction fixes the size of the top-layer commitments while allowing the data size to grow by adding layers. As a result, it achieves constant commitment size, but incurs a higher sampling cost for a given level of certainty due to stopping-set phenomena in LDPC codes under peeling decoding. \cite{LDPCForCMT} proposes a more careful LDPC code design for DAS by systematically eliminating stopping sets and optimizing the corresponding indexing technique.

\par
The undecodability ratio depends on both the code and employed decoder. For a 2D-RS code at rate $r=0.25$, up to $75\%$ of coded symbols can be revealed without enabling reconstruction, so $\alpha = 0.25$ \cite{2DRSDataAvailability}. The LDPC codes considered in \cite{LDPCForCMT, CodedMerkleTree}, decoded via peeling, exhibit $\alpha \approx 0.125$.

\subsubsection{Example paradigm 2: Sampling through RLNC}
\label{sec:RLNC_sampling}

\par
A possible way to implement sampling by coding leverages random linear network codes (RLNC). In this implementation, the verifier sends the claimer a vector consisting of coding coefficients for an RLNC coding operation.
The claimer node holds the data they claim to be available in the form of a matrix $\bm{V}\in \mathbb{F}^{m\times n}$
The claimer node responds to this challenge by sending an RLNC-coded data vector, obtained as a linear combination of the $n$ column vectors of the matrix $\bm{V}$ in its custody, using the specified coefficient vector. 
Since the decoding algorithm for RLNC is based on inverting the matrix of coefficient vectors, a claimer must render all coded vectors corresponding to one particular degree of freedom in the space of coefficient vectors unavailable. To make the data thus available, the claimer can only reveal a subspace of dimension at most $n-1$. Consequently, the claimer can answer at most $|\mathbb{F}|^{\,n-1}$ out of the $|\mathbb{F}|^n$ possible challenges. The corresponding undecodability ratio is calculated as
\begin{align*}
    \alpha_{\text{RLNC}} = 1 - \frac{1}{|\mathbb{F}|}.
\end{align*}

\par
This means that when sampling from RLNC with 1-byte coefficients, the uncdecodability ratio amounts to $1 - \tfrac{1}{256} \approx 99.6\%$ of the samples, while with 2-byte coefficients this increases to $1 - \tfrac{1}{65536} \approx 99.998\%$.

\par
Fig.~\ref{fig:prob_soundness_failure} compares the impact of applying two different paradigms in the context of DAS. The paradigm of sampling by indexing is represented through indexing coded data from LDPC and 2D-RS codes, with undecodability ratios of $\alpha = 0.25$ and $\alpha = 0.125$, respectively, as used in \cite{2DRSDataAvailability, CodedMerkleTree, PeerDAS}. The figure shows the probability that a verifier fails to detect a malicious claimer who withholds enough samples to make decoding impossible, after successfully retrieving $s$ samples. For a target failure probability of $10^{-4}$, the verifier must obtain 33 samples from the RS code and 70 samples from the LDPC code.

\par
By contrast, under the sampling from coding paradigm, exemplified through RLNC, a single sample suffices when the size of the field is bigger than two bytes. As we will see in the concrete instantiation of a DAS protocol with RLNC, these probabilities directly correspond to the probability of the soundness property being broken.

\section{RLNC-DAS}
\label{sec:protocol_description}

\subsection{Description}

\begin{figure}
    \centering
    \includegraphics[width=\linewidth]{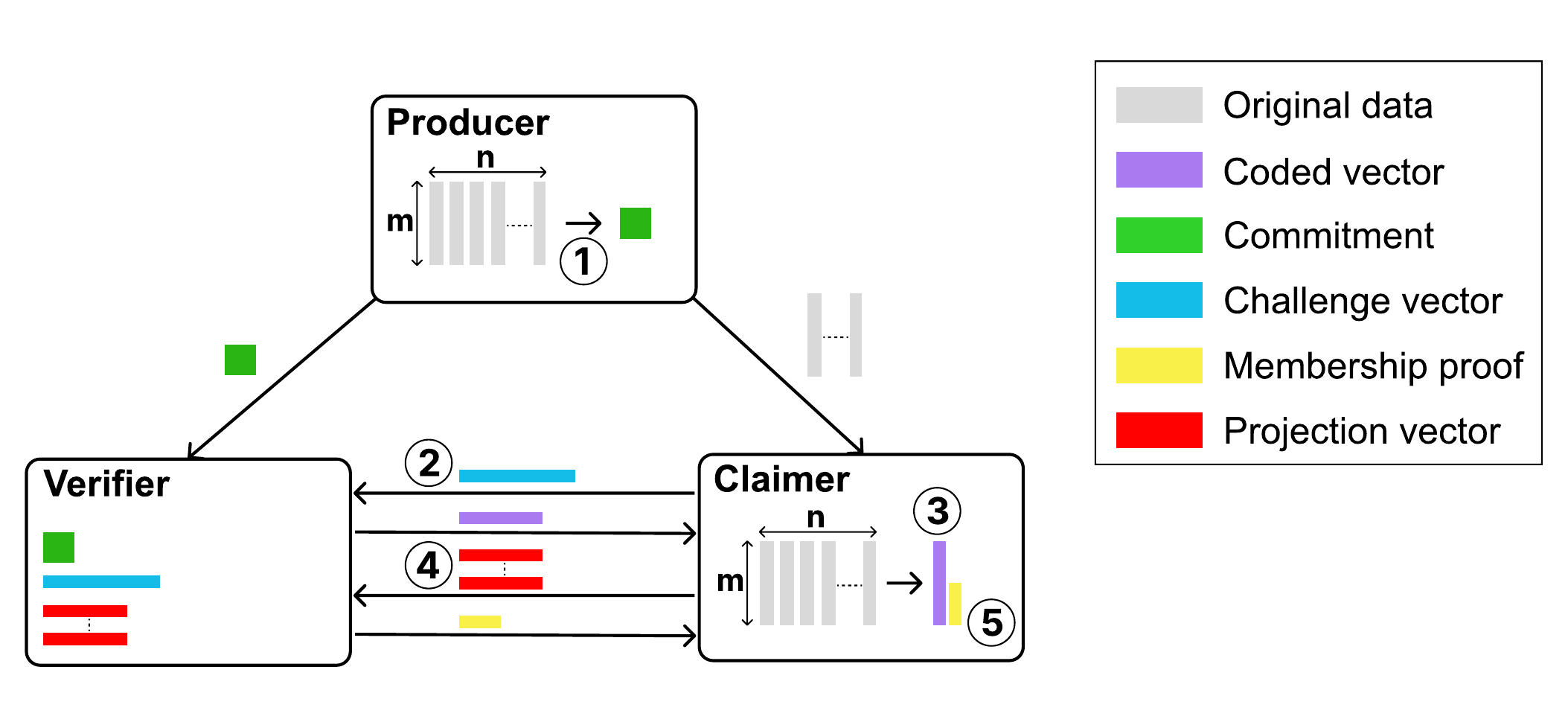}
    \caption{Information exchange between the system actors in the DAS protocol using RLNC-based on-the-fly sampling, together with a vector (Pedersen) commitment and an interactive inner product argument proof.}
    \label{fig:model_rlnc_DAS}
\end{figure}

\par
Here we show one possible way to design a DAS protocol using sampling through coding leveraging RLNC and a vector commitment on the row vectors of the uncoded data matrix $\bm{V}\in\mathbb{F}^{m\times n}$. To verify the consistency of the coded vector with the cryptographic commitment to the uncoded data matrix, we use a second step where the verifier requests the claimer to send an inner product argument \cite{bunz2018bulletproofs, bunz2021proofs} for the projection of the coded vector on vectors chosen by the verifier. The resulting scheme has favorable properties in terms of a constant commitment size, logarithmic download size in the total amount of data due to the inner product argument in the verification step and no storage overhead at the claimer node as samples are generated on the fly.

\par
The procedures that make up the protocol are described as follows, and shown in Fig. \ref{fig:model_rlnc_DAS}
\begin{enumerate}
    \item \textbf{Commitment generation}: Given a data matrix $\bm{V} \in \mathbb{F}^{m\times n}$ the data producer computes the commitments over rows with a Pedersen commitment: 
    \begin{align*}
        [[\bm{v}^{(i)}]]_{g} = \sum_{j=1}^{n}g_jV_{i,j} \; \in \mathbb{C}
    \end{align*}
    and publishes the commitments to the network.
    \item \textbf{Data availability challenge}: The verifier sends the claimer a vector of coefficients $\bm{c} \in \mathbb{F}^n$.
    \item \textbf{Challenge response}: Given the coefficients $\bm{c}$, the claimer sends the verifier the RLNC coded vector
    \begin{align*}
        \bm{\omega} = \sum_{i=1}^nc_i\bm{v}_i \; \in\mathbb{F}^m.
    \end{align*}
    \item \textbf{Membership proof request}: After receiving the coded vector $\bm{\omega}$, the verifier sends $p$ vectors $\{\bm{p}_i\in\mathbb{F}^m\}_{i=1}^{p}$ to the claimer.
    \item \textbf{Membership proof generation}: After receiving the vectors $\bm{p}_i$, the verifier sends a membership proof in the form of $p$ inner product arguments $\left<\left<\sum_{j=1}^mp_{i,j}\bm{v}^{(j)}, \bm{c}\right>\right>$. Each inner product argument contains $2\log(n)$ EC points and $2$ scalars in $\mathbb{F}$ 
    \item \textbf{Membership proof verification}: The verifier confirms the inner product arguments using
    \begin{itemize}
        \item Vector commitments: $[[\bm{c}]]$ and $\sum_{j=1}^{m}p_{i,j}[[\bm{v}^{(j)}]]$
        \item Inner product value: $\bm{p}_i^\top\bm{\omega}$
    \end{itemize}
    If even one verification fails, then the data is deemed unavailable. If it succeeds, then the verifier sends another challenge if it needs more certainty about data being available (step 2), else the verifier concludes that the data is available to the network
\end{enumerate}

\subsection{Security Properties}

\subsubsection{Definitions}

\par
There exist several security primitives for DAS schemes \cite{hall2023foundations, DAMultiplicity} that make different assumptions about the underlying system of nodes participating in the network as well as the power of a potential malicious claimer node.

\par
\textbf{Completeness}: Given an honest claimer answers the verifiers sampling request truthfully, the verifier will conclude that the data is available. This property captures the protocol’s correct behavior under the assumption that all actors act honestly.

\par
\textbf{Soundness}: If an honest verifier accepts the data to be available, then some data entity can be reconstructed with high probability.

\par
\textbf{Consistency}: For a set of coded data vectors received from the claimer node that satisfy soundness, that is, they enable the reconstruction of a data entity, the reconstructed entity coincides, with high probability, with the data that the producer committed to. This property guarantees that the reconstructible data remains bound to the original uncoded data that was committed. We say that a sample is consistent if it is truthfully derived from the underlying committed data entity.

\par
We now state theorems concerning the soundness and consistency properties of the protocol under the strong model for data availability. We do not state completeness as a separate theorem, since it follows directly from the protocol construction. Throughout, we assume a single field $\mathbb{F}$ for the projection, data, and coefficient vector entries, and denote its cardinality by $q$.

\subsubsection{Analysis}

\par
\textbf{Soundness}:
For the scheme to be sound, we have to check that if the data is deemed available by the verifier, then some data entity $\bm{V}\in\mathbb{F}^{m\times n}$ can be reconstructed from a sufficient amount of coded vectors sampled from the claimer node with high probability. Appendix \ref{appendix:soundness_proof} proves the following theorem: 
\begin{theorem}
\label{thrm:soundness}
    Assuming $l$ verifiers sample $s$ coded vectors each from the claimer (given $ls \geq n + s$), RLNC DAS satisfies soundness with probability at most $1/q^s$. 
\end{theorem} 
From this theorem we can make two observations:  
\begin{enumerate}
        \item Each verifier implicitly assume that a sufficient number of other verifiers are sampling coded data from the claimer node(s) corresponding to the same commitment. This is a natural condition for availability, since data can only be considered truly available to the network if enough coded pieces have been emitted by the claimer node and can subsequently be reconstructed across verifier nodes.
        \item The probability of soundness failure is the same as the probability of a verifier failing to detect that a malicious claimer only selectively responds to the data availability challenges, as described in Sec.~\ref{sec:RLNC_sampling}. This implies that soundness failure can be made sufficiently small with only a few samples compared to sampling from fixed-rate redundancy codes.
\end{enumerate}

\par
\textbf{Consistency}: 
For the scheme to be consistent, the probability that a verifier accepts an inconsistent sample from a claimer must be negligible. The following theorem, proved in Sec.~\ref{sec:appendix:consistency_proof}, formalizes this requirement:
\begin{theorem}
\label{thrm:consistency}
    Let a verifier send $p$ vectors $\{\bm{p}_i\}_{i=1}^p$ to the claimer as part of the membership proof. Then a malicious claimer can violate the consistency of a single coded vector with probability at most $1/q^p$.
\end{theorem}

We observe that the probability of a consistency failure decreases exponentially with the number of vectors $p$ sent by the verifier. Consequently, even a small number of vectors suffices to make the failure probability negligible. For example, for a field size of $q = 2^{256}$ (corresponding to 32-byte symbols) and $p = 1$, the failure probability is $2^{-256}$, so the claimer must succeed on this event in a single attempt.

\section{Numerical Analysis}

\begin{table}[t]
\renewcommand{\arraystretch}{1.15}
\centering
\caption{Simulation parameters for DAS protocols.}
\begin{tabular}{|l|l|l|}
\hline
\textbf{Parameter} & \textbf{Value} & \textbf{Description} \\
\hline
\multicolumn{3}{|c|}{\bfseries RLNC-DAS} \\
\hline
$|\mathbb{F}|$ & $32\text{B}$  & Size of coefficients and data chunks \\
$|\mathbb{C}|$ & $48\text{B}$ & Size of elliptic-curve group elements \\
$p$            & 1         & Number of vectors to verify membership \\
\hline
\multicolumn{3}{|c|}{\bfseries LDPC-based Coded Merkle Tree} \\
\hline
$b$        & 256B  & Size of base-layer symbol in coded Merkle tree \\
$r$        & 0.25  & Coding rate: ratio of encoded symbols to information symbols \\
$\alpha$   & 0.125 & Undecodability ratio \\
$q$        & 8     & Batching factor: symbols concatenated per hash at each level \\
$t$        & 256   & Number of root hashes \\
\hline
\multicolumn{3}{|c|}{\bfseries 2D-RS \cite{2DRSDataAvailability}} \\
\hline
$b$ & 512B & Size of one (un)coded symbol \\
$r$ & 0.25 & Coding rate: ratio of encoded symbols to information symbols \\
\hline
\end{tabular}
\label{tab:sim-params}

\end{table}

\par
In this section, we evaluate RLNC-DAS, as described in Sec.~\ref{sec:protocol_description}, with respect to its overall bandwidth cost, focusing primarily on the interaction between claimer and verifier and, in particular, on the verifier's download cost. We further provide asymptotic evaluations of the computational complexity incurred during commitment, coding, and membership proving for a several verifier sampling from one claimer. For this purpose, we compare RLNC-DAS against three data availability sampling (DAS) protocols from the literature, considered in their originally proposed or slightly adapted form:

\begin{enumerate}
    \item Sampling from a coded Merkle tree \cite{CodedMerkleTree},
    \item Two protocols based on 2D-RS codes, using either Merkle trees or KZG commitments as the underlying commitment scheme \cite{2DRSDataAvailability}.
\end{enumerate}

The parameters used for this comparison are listed in Table~\ref{tab:sim-params}. To the best of our knowledge, they reflect realistic and widely adopted design choices for DAS schemes that have either been deployed in practice or proposed in the literature. Following common design choices in blockchains such as Ethereum, we instantiate the Pedersen commitment over the BLS12-381 curve. This curve provides compressed group elements of size $48$B and a prime group order of approximately $32$B. We adopt the latter as the chunk size for the RLNC data vectors to ensure compatibility with the linear homomorphism of the Pedersen commitments.

The RLNC-DAS construction achieves a consistency failure probability of $2^{-256}$, corresponding to a 256-bit consistency level. This aligns with the security targets commonly adopted for cryptographic primitives in blockchain systems.

\subsection{Performance Evaluation}

\begin{table*}[t]
  \centering
  \caption{Comparison of DAS schemes for 32\,MB of data with $P_{f,\mathrm{soundness}}\leq10^{-9}$.}
  \label{tab:32MB_comparison}
  \renewcommand{\arraystretch}{1.15}
  \begin{tabular}{|l|l|c|c|c|c|c|}
    \hline
    \textbf{Code} &
    \makecell{\textbf{Commit-}\\\textbf{ment}} &
    \makecell{\textbf{Sampling}\\\textbf{cost}} &
    \makecell{\textbf{Comm.}\\\textbf{size}} &
    \makecell{\textbf{Storage}\\\textbf{overhead}} &
    \makecell{\textbf{Auditor}\\\textbf{node}} &
    \makecell{\textbf{Trusted}\\\textbf{setup}} \\
    \hline
    2D-RS & MT        & 57.0\,kB    & 32.0\,kB & 96.0\,MB & Yes & No  \\
    2D-RS & KZG       & 38.8\,kB  & 8.0\,kB  & 96.0\,MB & No  & Yes \\
    LDPC & CMT       & 736.1\,kB & 8.0\,kB  & 96.0\,MB & Yes & No  \\
    RLNC ($m=16$) & Pedersen & 3.0\,kB   & 0.8\,kB  & 0.0\,B   & No  & No  \\
    \hline
  \end{tabular}
\end{table*}

\begin{figure*}[t]
  \centering
  \pgfplotslegendfromname{da-unified-legend}\vspace{0.6em}

  \begin{tikzpicture}
    \begin{groupplot}[
      group style={group size=2 by 1, horizontal sep=2cm},
      width=0.49\textwidth,
      height=6cm,
      xmode=log, log basis x=2,
      ymode=log,
      xtick={1,4,16,64,256,1024},
      xticklabels={1,4,16,64,256,1024},
      minor x tick num=0,
      ytick={1e-6,1e-5,1e-4,1e-3,1e-2},
      minor y tick num=0,
      grid=major, major grid style={gray!30,dashed},
      label style={font=\large}
    ]

      \nextgroupplot[
        xlabel={d in MB},
        ylabel={Sampling size / d},
        legend to name=da-unified-legend,
        legend columns=4,
        legend style={draw=black, fill=white, rounded corners=2pt}
      ]
        \addplot[rlnc, color=rb1] table[x=d, y=rlnc16, col sep=comma]{\DARESP};   \addlegendentry{RLNC (m=16)}
        \addplot[rlnc, color=rb3] table[x=d, y=rlnc64, col sep=comma]{\DARESP};  \addlegendentry{RLNC (m=64)}
        \addplot[rlnc, color=rb4] table[x=d, y=rlnc256, col sep=comma]{\DARESP};  \addlegendentry{RLNC (m=256)}
        \addplot[rlnc, color=rb5] table[x=d, y=rlnc1024, col sep=comma]{\DARESP}; \addlegendentry{RLNC (m=1024)}
        \addplot[twodrs, mark=|] table[x=d, y=twodrs_mt, col sep=comma]{\DARESP}; \addlegendentry{2D-RS + MT}
        \addplot[twodrs] table[x=d, y=twodrs_kzg, col sep=comma]{\DARESP}; \addlegendentry{2D-RS + KZG}
        \addplot[cmt] table[x=d, y=cmt, col sep=comma]{\DARESP}; \addlegendentry{CMT}

      \nextgroupplot[
        xlabel={d in MB},
        ylabel={Commitment size / d}      ]
        \addplot[rlnc, color=rb1] table[x=d, y=rlnc16, col sep=comma]{\DACOMM};
        \addplot[rlnc, color=rb3] table[x=d, y=rlnc64, col sep=comma]{\DACOMM};
        \addplot[rlnc, color=rb4] table[x=d, y=rlnc256, col sep=comma]{\DACOMM};
        \addplot[rlnc, color=rb5] table[x=d, y=rlnc1024, col sep=comma]{\DACOMM};
        \addplot[twodrs, mark=|] table[x=d, y=twodrs_mt, col sep=comma]{\DACOMM};
        \addplot[twodrs] table[x=d, y=twodrs_kzg, col sep=comma]{\DACOMM};
        \addplot[cmt] table[x=d, y=cmt, col sep=comma]{\DACOMM};

    \end{groupplot}
  \end{tikzpicture}

  \caption{Comparison of verifier download costs relative to the total amount of data d for RLNC-coded DAS under different choices of coded vector length $m$, sampling by indexing DAS protocols for a target probability of soundness error of $10^{-9}$. The latter combine either layered approaches (CMT, Merkle Trees) or polynomial commitment schemes (KZG) with fixed-rate redundancy codes (2D-RS / LDPC in the case of CMT). \textbf{Left:} DA challenge response size, consisting of a sampled code symbol together with its membership proof, shown relative to the total data size $d$. \textbf{Right:} Commitment size, shown relative to the total data size $d$ for these schemes.
}
  \label{fig:da-verifier_download_cost}
\end{figure*}

\subsubsection{Light Node Download Cost}

\par
Fig. \ref{fig:da-verifier_download_cost} compares the light node download cost of RLNC DAS to the other protocols in terms of commitment and sampling cost. Four instantiations of RLNC DAS with different vector lengths ($m\in\{16, 64, 256, 1024\}$) are displayed. The data size is itself grown with the number of data vectors $n$ included. For all schemes, we aim for a target soundness failure rate of $< 10^{-9}$.

\par
\begin{itemize}

    \item \textbf{Sampling cost}: For RLNC, one DA challenge has to be answered correctly by the claimer to convince the verifier with a certainty of $\geq1-10^{-9}$ that the data is indeed available. The entire download cost thus sums up to one coded vector with a size of $m|\mathbb{C}|$ and one inner product argument over two length $n$ vectors with a total size of $2(\log_2(n)|\mathbb{C}| + |\mathbb{F}|)$ each. For the CMT and 2D-RS constructions, on the other hand,  73 and 156 DA challenges have to all be answered respectively. This results in up to $\sim 10$ times lower sampling costs compared to the DAS schemes using 2D-RS codes, and $\sim 200$ compared to sampling from a coded Merkle tree when choosing $m=16$.

    \item \textbf{Commitment size}: With RLNC DAS, the commitment size stays constant with $m\cdot|\mathbb{C}|$ and can thus be controlled in its size by choosing the dimension of the data vector $m$. The CMT also provides a constant size commitment by construction ($8kB$ in this case). 2D-RS paired with Merkle trees creates a commitment for every row and every column, which is why the commitment size scales with the square root of the total amount of data $d$.
\end{itemize}

\par
Table~\ref{tab:32MB_comparison} presents the download cost comparison for 32~MB of total data. Among the listed protocols, the combination of RLNC with Pedersen commitments is the only one that requires neither a trusted setup nor the presence of an honest auditor node. Furthermore, for $m=16$, RLNC-DAS achieves a total download cost of $3.8\text{kB}$ for the verifier node, reducing the cost by a factor of $\sim 12$ compared to 2D-RS+KZG and by a factor of $\sim195$ compared to CMT+LDPC~\cite{CodedMerkleTree}. Another advantage of RLNC-DAS is that there is no storage overhead at the claimer nodes, since samples are generated on demand, at the expense of additional computational complexity during sampling. In contrast, redundancy-coded schemes based on RS and LDPC codes with rate $r=0.25$ require storing the redundancy-coded data, forcing claimers to store at least $4\times$ the original data (an additional $96\text{MB}$ in this case).

\par
These results illustrate that, as an embodiment of the sampling by coding paradigm, as RLNC-DAS requires fewer samples to achieve the same soundness error, this directly translates to a reduction in sampling cost for light nodes compared to schemes based on the sampling by indexing paradigm. The difference is particularly recognizable for fixed-rate redundancy codes with low undecodability ratios, exemplified through a CMT employing LDPC codes, whose stopping sets yield an undecodability ratio of $\alpha_{\mathrm{LDPC}} = 0.125$. Moreover, the flexibility of RLNC allows the commitment size to be controlled by the system parameter $m$. In contrast, LDPC-based constructions such as CMT, and 2D-RS-based constructions, do not offer comparable flexibility due to stopping-set and field-length constraints, respectively.

\begin{table*}[t]
  \centering
  \caption{Comparison of total networking cost from the view of a DA verifier of DAS schemes for 32\,MB of data with $P_{f,\mathrm{soundness}}\leq10^{-9}$.}
  \label{tab:networking_cost}
  \renewcommand{\arraystretch}{1.15}
  \begin{tabular}{|l|l|c|c|c|}
    \hline
    \textbf{Code} &
    \makecell{\textbf{Commit-}\\\textbf{ment}} &
    \makecell{\textbf{Upload cost}} &
    \makecell{\textbf{Download cost}} &
    \makecell{\textbf{Total network}\\\textbf{cost}} \\
    \hline
    2D-RS & MT        & $288.0$\,B & 89.0\,kB & 89.0\,kB   \\
    2D-RS & KZG       & $288.0$\,B & 46.8\,kB & 46.8\,kB  \\
    LDPC  & CMT       & $48.8$\,kB & $744.1$\,kB & $792.9$\,kB  \\
    RLNC ($m=16$) & Pedersen & $64.0$\,B & $3.8$\,kB & $3.8$\,kB  \\
    \hline
  \end{tabular}
\end{table*}

\subsubsection{Total Bandwidth Cost}
Table~\ref{tab:networking_cost} evaluates the verifier upload cost and the total network cost incurred by the different protocols, including the number of interactions required per sampling operation, for $d = 32$\,MB of data, extending the total verifier download cost from table \ref{tab:32MB_comparison}.

\begin{enumerate}
    \item RLNC-DAS requires the verifier to send one vector $\bm{p\in\mathbb{F}^m}$ and one coefficient vector $\bm{c}\in\mathbb{F}^n$ to the claimer, which together constitute the data availability challenge.
    \item 2D-RS sampling requires transmitting only two indices to the claimer.
    \item For the coded Merkle tree (CMT), the verifier samples one element from each layer of the tree and, with probability $1-r$, additionally samples from each intermediate layer.
\end{enumerate}

While cursory considerations suggest that the coefficient vectors in RLNC-DAS scale with the total data size and thus make the verifier upload cost a dominant component of the overall network cost, the vectors $p$ and $n$ in RLNC-DAS merely serve to define a random challenge that must remain unpredictable at the time the data is committed. Rather than transmitting the full vectors explicitly, the verifier can instead send short seed values for mutually agreed-upon cryptographic hash functions. The full challenge vectors can then be derived deterministically from these seeds, substantially reducing the upload cost while preserving the soundness guarantees. This optimization is conceptually aligned with established RLNC techniques for reducing protocol overhead~\cite{heide2017rlnc}.

\subsubsection{Computational Cost}

\begin{table*}[t]
  \centering
  \caption{Asymptotic computational cost of DA claimer in DAS protocols in a setting with $l$ verifiers requesting $s$ samples each. We distinguish between sampling, proving, and commitment cost. Here, 2D-RS uses code dimensions $(\sqrt{n},\sqrt{k})$ in each dimension and RLNC-DAS leverages an $m \times n$ data matrix   .}
  \label{tab:computational_cost}
  \renewcommand{\arraystretch}{1.15}
  \begin{tabular}{|l|l|c|c|c|}
    \hline
    \textbf{Code} &
    \textbf{Commitment} &
    \makecell{\textbf{Coding} \\ \textbf{cost}} &
    \makecell{\textbf{Proving} \\ \textbf{cost}} &
    \makecell{\textbf{Commitment} \\ \textbf{cost}} \\
    \hline
    2D-RS & MT  & $\mathcal{O}(n\log(n))$ & $\mathcal{O}(1)$ & $\mathcal{O}(n)$ \\
    2D-RS & KZG & $\mathcal{O}(n\log(n))$ & $\mathcal{O}(n^{3/2})$ & $\mathcal{O}(k)$ \\
    LDPC & CMT & $\mathcal{O}(n^2)$ & $\mathcal{O}(1)$ & $\mathcal{O}(n^2)$ \\
    RLNC-DAS & Pedersen & $\mathcal{O}(mnls)$ & $\mathcal{O}(lsn)$ & $\mathcal{O}(mn)$ \\
    \hline
  \end{tabular}
\end{table*}

Lastly, we compare the computational complexity of RS- and LDPC-based sampling methods with RLNC-based sampling in Table~\ref{tab:computational_cost}. The comparison distinguishes between three components: coding, proving, and commitment generation. At a high level, the key difference is that RS- and LDPC-based approaches follow a sampling-by-indexing paradigm, where the claimer first computes a coded version of the data and then answers verifier queries by revealing precomputed coded symbols together with the corresponding authentication information. In contrast, RLNC-based sampling follows a sampling-by-coding paradigm, where each verifier request is answered by generating a fresh coded sample and its associated proof on demand, requiring additional computation upon verifier request.

For the RS-based schemes, the coding cost is therefore a one-time preprocessing cost that depends only on the code parameters $n, k$. In the Merkle-tree-based variant, the commitment cost contains constructing the authentication structure over the coded data, i.e., computing the leaf hashes and internal hashes needed to build the Merkle trees for the relevant rows and columns. The corresponding proofs are then available as authentication paths in the committed structure, so they do not require additional computation beyond selecting the appropriate path for a queried symbol. In the KZG-based variant, commitment generation requires computing one commitment per row, while proving requires the explicit computation of KZG opening proofs for the requested coded chunks.

For the LDPC-based scheme, the coding cost is likewise incurred as a preprocessing step and depends only on the chosen code parameters. Commitment generation is performed through the construction of the coded Merkle tree. In contrast to a standard Merkle tree, this requires not only hashing and concatenation at each layer, but also re-encoding the symbols of each layer before proceeding to the next one. Since the number of symbols decreases geometrically across layers, the total commitment cost remains asymptotically dominated by the bottom layer. As in the Merkle-tree-based RS construction, the corresponding proofs are already contained in the committed structure and can therefore be served without additional expensive prover-side computation.

RLNC-DAS differs in that coding and proving are performed in response to verifier queries. For each requested sample, the claimer computes a new random linear combination of the coded data and then generates the corresponding proof of membership or consistency. As a result, its computational cost scales with the number of verifier requests $ls$, captured here by the number of verifiers and the number of samples requested by each verifier.

Overall, RS- and LDPC-based sampling methods incur a fixed upfront cost for encoding and commitment generation, after which samples can be served efficiently. RLNC-DAS, on the other hand, only creates as many samples as requested. Consequently, RLNC-DAS is computationally attractive from the claimer perspective when the number of verifier requests is small, a benefit further reinforced by its higher sampling efficiency. In contrast, RS- and LDPC-based methods become more favorable in settings with many verifiers or with many requested samples per claimer.

\section{Discussion}

\par
We presented a new approach to designing DAS protocols, where sampling is performed through on-the-fly coding rather than by indexing pre-committed redundancy coded data. The advantages of this new approach were demonstrated through a concrete protocol instantiation based on RLNC. The resulting scheme exhibits desirable properties, particularly in terms of storage efficiency for the claimer node and reduced bandwidth requirements for the verifier node.

\par
Further work could extend the security analysis presented here to stronger adversarial models \cite{DAMultiplicity, hall2023foundations}.

\par
Although the protocol described here assumes that claimer nodes hold the full uncoded data, the sampling by coding paradigm is not restricted to that setting. A claimer node may instead prove properties about a partial data representation it possesses, or multiple claimer nodes may collaborate to jointly respond to a verifier’s DA challenge. Once again, RLNC is a natural fit because of its composability and repairability \cite{deb2005good, abdrashitov2015durable}: a node holding only partial coded data can still generate fresh, valid codewords by recoding coded data.

\bibliographystyle{splncs04} 
\bibliography{references}

\appendix

\section{Proof of Soundness Theorem}
\label{appendix:soundness_proof}

\par
For soundness, some data entity must be reconstructible from the emitted coded vectors whenever the verifier accepts data availability with high probability. We prove this theorem for the strong model for data availability \cite{neu2022data}. We introduce the following notation:
\begin{itemize}
    \item $\mathcal{H}$: the event that the claimer is honest, i.e., responds to all requests $\bm{c}$ with a coded vector $\bm{\omega}$.
    \item $\mathcal{V}$: the event that the verifier accepts that the data is available (some data can be reconstructed from the coded vectors).
    \item $q$: the field size, $|\mathbb{F}|$.
    \item $\bm{C}=[\bm{c}_1, \dots, \bm{c}_{ls}] \in \mathbb{F}^{n \times ls}$: the matrix whose columns are the coefficient vectors of sampled coded vectors $\bm{\omega}_i$.
\end{itemize}

\noindent\textbf{Theorem}
\textit{Assuming $l$ verifiers sample $s$ coded vectors each from the claimer (given $ls \ge n+s$), RLNC DAS satisfies soundness with probability at most $1/q^s$. }

\begin{proof}
Since reconstruction from $ls$ RLNC coded vectors uses Gaussian elimination, soundness fails exactly when $\operatorname{rank}(\bm{C})<n$. Decomposing over $\mathcal{H}$ and $\bar{\mathcal{H}}$ gives
\begin{align}
P\bigl(\mathcal{V} \cap \{\operatorname{rank}(\bm{C})<n\}\bigr)
&= P\bigl(\mathcal{H}\bigr)\, P\bigl(\mathcal{V} \cap \{\operatorname{rank}(\bm{C})<n\}\mid \mathcal{H}\bigr) \notag\\
&\quad + P\bigl(\bar{\mathcal{H}}\bigr)\, P\bigl(\mathcal{V} \cap\{\operatorname{rank}(\bm{C})<n\}\mid \bar{\mathcal{H}}\bigr) \notag\\
&\le \max\!\left\{
P\bigl(\mathcal{V} \cap \{\operatorname{rank}(\bm{C})<n\}\mid \mathcal{H}\bigr),\;
P\bigl(\mathcal{V} \cap \{\operatorname{rank}(\bm{C})<n\}\mid \bar{\mathcal{H}}\bigr)
\right\}.
\label{eq:decomp}
\end{align}

Under non-adaptivity, the probability that the verifier accepts in the dishonest case is
\begin{align}
    P\bigl(\mathcal{V} \cap \{\operatorname{rank}(\bm{C})<n\}\mid \bar{\mathcal{H}}\bigr) = \frac{1}{q^s}.
    \label{eq:bad}
\end{align}

By completeness, $\mathcal{H} \Rightarrow \mathcal{V}$, hence
\begin{align*}
    P\bigl(\mathcal{V} \cap \{\operatorname{rank}(\bm{C})<n\}\mid {\mathcal{H}}\bigr)
    = P\bigl(\{\operatorname{rank}(\bm{C})<n\}\mid {\mathcal{H}}\bigr).
\end{align*}
A counting argument over $\bm{C}\in\mathbb{F}^{n\times ls}$ \cite{networkCodingForEngineers} yields
\begin{align*}
    P\bigl(\{\operatorname{rank}(\bm{C})<n\}\mid {\mathcal{H}}\bigr)
    &= 1 - P\bigl(\{\operatorname{rank}(\bm{C})=n\}\mid {\mathcal{H}}\bigr) \notag\\
    &= 1 - \dfrac{\prod_{i=0}^{n-1}(q^{ls} - q^i)}{q^{lsn}}
     \;=\; 1 - \prod_{i=0}^{n-1}(1-q^{i-ls}).
\end{align*}

Using the inequality
\begin{align*}
    \prod_{i=0}^{n-1}(1-q^{i-ls})  \geq 1-\sum_{i=0}^{n-1}q^{i-ls},
\end{align*}
which can be proved by induction, we obtain the bound
\begin{align}
    P\bigl(\{\operatorname{rank}(\bm{C})<n\}\mid {\mathcal{H}}\bigr)
    \leq \sum_{i=0}^{n-1}q^{i-ls}
    = q^{-ls}\dfrac{q^{n}-1}{q-1}
    \le q^{n-ls}.
    \label{eq:geom}
\end{align}

Combining \eqref{eq:decomp}, \eqref{eq:bad}, and \eqref{eq:geom}, and noting that for $ls\geq n + s$ the bound in \eqref{eq:geom} is smaller than or equal to $1/q^s$, we conclude that the soundness-failure probability bound is
\begin{align*}
 P\bigl(\mathcal{V} \cap \{\operatorname{rank}(\bm{C})<n\}\bigr) \le \dfrac{1}{q^s}.
 \label{eq:final}
\end{align*}
\end{proof}

\section{Proof of Consistency Theorem}
\label{sec:appendix:consistency_proof}

\par
For consistency, any data sample accepted by the verifier must be drawn from the same data matrix on which the commitment was formed. For readability, we use the following notation:
\begin{itemize}
    \item $q$: field size $|\mathbb{F}|$.
    \item $p$: number of vectors used for membership verification, $\{\bm{p}_i\}_{i=1}^p$.
    \item $\mathcal{C}$: event that the claimer breaks consistency of a sample, that is, makes an honest verifier accept $\bm{\omega}'\neq \bm{V}\bm c$ for a data availability challenge $\bm c$
\end{itemize}
Further, we make the following assumption:
\begin{itemize}
    \item An adversary cannot break the inner product argument; that is, they cannot construct an inner product argument $<<\bm{a}, \bm{b}>>$ that opens to values different from $\bm a$, $\bm b$, and $\bm a^{\top}\bm b$.
\end{itemize}

\noindent\textbf{Theorem.} 
\textit{Let a verifier send $p$ vectors $\{\bm{p}_i\}_{i=1}^p$ to the claimer as part of the membership proof. 
Then a malicious claimer can violate the consistency of a single coded vector with probability at most $1/q^p$.}

\begin{proof}
Under the assumption that the inner product argument is unbreakable, for the verifier to accept the claimer’s response, a malicious claimer must send a vector $\bm{\omega}' = \bm{V}\bm{c} + \bm{d}$ with $\bm{d}\neq\bm{0}\in\mathbb{F}^m$ such that it satisfies the same inner-product relations with $\{ \bm{p}_i \}_{i=1}^{p}$ as $\bm{\omega} = \bm{V}\bm{c}$; i.e., $\bm{p}_i^{\top}\bm{\omega}' = \bm{p}_i^{\top}\bm{\omega}$ for all $i$. Therefore,
\begin{align*}
    P(\mathcal{C})
    &= P\!\left(\left\{ \bm{p}_i^{\top}\bm{\omega}' = \bm{p}_i^{\top}\bm{\omega} \;\; \forall i \in \{1,\dots,p\} \right\}\right) \\
    &= P\!\left(\left\{ \bm{p}_i^{\top}\bm{d} = \bm{0} \;\; \forall i \in \{1,\dots,p\} \right\}\right).
\end{align*}

Since $\bm{d}$ and $\bm{p}_i$ are chosen independently, the quantities $\bm{p}_i^{\top}\bm{d}$ are i.i.d. and uniformly distributed over $\mathbb{F}$. Consequently,
\begin{align*}
     P(\mathcal{C}) &= \prod_{i=1}^p P(\{\bm{p}_i^{\top}\bm{d}=\bm{0}\}) = \dfrac{1}{q^{p}}.
\end{align*}
\end{proof}

\end{document}